Magnetic hard nanobubble: a possible magnetization structure behind the bi-skyrmion

Yuan Yao (姚湲)[1,†,*], Bei Ding (丁贝)[1, 2,†], Jie Cui (崔婕)[1, 2], Xi Shen(沈希)[1], Yanguo Wang (王岩国)[1], Wenhong Wang (王文洪)[1], Richeng Yu (禹日成)[1, 2]

1. Beijing National Laboratory of Condensed Matter Physics,

Institute of Physics, Chinese Academy of Sciences,

Beijing 100190, China

2. School of Physical Sciences, University of Chinese Academy of Science

Beijing 100049, China

† Equal contribution for this work.

* Corresponding author, email: yaoyuan@iphy.ac.cn

Abstract

Transport of intensity equation (TIE) has been applied to process the simulated and experimental images of the magnetic hard nanobubbles, which were acquired in the Lorentz transmission electron microscope (LTEM). Systematic studies demonstrated that the processing parameter in TIE can modulate the features of the retrieved magnetization and induce the bi-spiral structures which may be identified as the bi-skyrmions.

With a definite chiral, the skyrmion is one of the research focuses in the magnetic materials community recently since it is considered as a promising candidate for high density memories and other modern electronic devices.[1, 2] Lorentz transmission electron microscope (LTEM) is a powerful apparatus to characterize the magnetic structures at the nanometer scale. With the aid of a certain image post-processing method, such as transport of intensity equation (TIE), magnetization distribution in the materials can be disclosed from the LTEM images. TIE method is an implement to retrieve the phase $\varphi(x,y)$ of the exit wave in electron microscopy with several images acquired at different defocuses,[3]

$$\varphi\left(x, y, z_0\right) = -k \nabla_{x,y}^{-2}\left\{\nabla_{x,y}\left[\frac{\nabla_{x,y}\nabla_{x,y}^{-2}\left(\frac{\partial I\left(x, y, z\right)}{\partial z}\right)}{I\left(x, y, z_0\right)}\right]\right\} \tag{1}$$

where $z$ is the electron propagation direction and $k$ is the electron wave vector. The partial differential of intensity $\partial I/\partial z$ means the intensity variation of images along $z$, which can be substituted by numerical differential between the image intensity at different focus $\Delta I/\Delta z$. Under the assistance with Fourier transfer (FT), the inverse Laplacian $\nabla_{x,y}^{-2}$ could be replaced by[4]

$$\nabla_{x,y}^{-2} f\left(x,y\right) = \mathcal{F}^{-1}\left\{\frac{\mathcal{F}\left[f\left(x,y\right)\right]}{|q\left(x,y\right)|^2}\right\} \tag{2}$$

where $q(x,y)$ is the spatial frequency in the image plane. However, eq. 2 is divergent and the noise is amplified when $q(x,y)$ approaching zero. So a small nonzero constant $q_0$ (or a modified Tikhonov-type filter) must be appended to avoid the divergence and suppress low-frequency disturbance (mainly the diffraction contrast) in the actual application[4, 5]

$$\nabla_{x,y}^{-2} f\,'\left(x,y\right) = \mathcal{F}^{-1}\left\{\frac{\mathcal{F}\left[f\left(x,y\right)\right]}{|q\left(x,y\right)|^2 + q_0^{\;2}}\right\} \tag{2'}$$

The magnetization in the image plane could be deduced from the partial differential of the obtained phase $\varphi(x,y)$[6]

$$M_x^{'} \propto B_x^{'} \propto \frac{\partial\varphi(x,y)}{\partial y}, \quad M_y^{'} \propto B_y^{'} \propto -\frac{\partial\varphi(x,y)}{\partial x} \tag{3}$$

It should be emphasized that $M_x'$ and $M_y'$ (or $B_x'$ and $B_y'$) are not the actual magnetization components (or magnetic induction components) because the retrieved exit wave phase is the line integral of the vector potential $\vec{A}(x,y,z)$ along the electron path $z$,

$$\varphi(x,y) = -\frac{e}{\hbar}\int_{-\infty}^{+\infty}\vec{A}(x,y,z)\cdot d\vec{z} = -\frac{e}{\hbar}\int_{-\infty}^{+\infty}A_z(x,y)dz \tag{4}$$

if the electron potential in specimen can be ignored.

TIE using FT approach solves the phase problem easily and is convenient to deal with the LTEM images. Various skyrmions have been revealed with LTEM, booming the research development in magnetic skyrmions.[7] Besides the skyrmions with unique $\pm1$ chirality, a special bi-skyrmion containing two opposite swirling contours is also an interesting magnetic configuration.[8, 9] Contrary to those novel $\pm1$ skyrmions, the formation mechanism of the bi-skyrmions with the skyrmion number (Ns) of $2$[8] is still unclear. Takagi et al. suggested that Dzyaloshinskii-Moriya interaction (DMI) should generate the bi-skyrmion arrays in non-centrosymmetric $Cr_{11}Ge_{19}$.[10] But in some centrosymmetric compounds, such as $La_{2-2x}Sr_{1+2x}Mn_2O_7$ [8, 11] and NiMnGa alloy [9, 12], the nature of the bi-skyrmion is still waiting to be illustrated. In our previous work, a simple Neel type magnetic nanodisk can show the bi-chiral structure in the recovered LTEM images when its symmetric axis deviates from the incident electron beam.[13] Such artifact comes from the projection characteristics of imaging in LTEM (Eq. 4), which may disappear when the axis is along the electron beam. In this report, another common magnetic nanostructure, magnetic hard nanobubble, is characterized systematically in both simulated and experimental images in LTEM. The results demonstrate that the bi-skyrmions may be reproduced in the retrieved images of the magnetic hard nanobubble if an improper parameter is selected in TIE processing, no matter how the specimen is placed in LTEM.

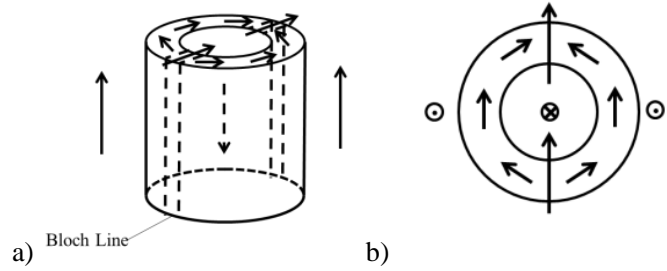

Fig. 1 a) Schematic configuration of the magnetization in a hard bubble, the arrows denote the magnetization orientation. b) The top view.

A magnetic hard bubble (HB) is a mixture of Bloch and Neel type domain walls. It is also defined as type II magnetic bubble because of its zero chiral number, while a bubble with the pure closed Bloch wall is named simple or soft bubble (SB) with a nonzero chirality.[14] Fig.1 is the schematic picture of an isolated HB where two semi-circle Bloch domain walls are separated by two Neel type gaps. The gaps are sometimes denoted as Bloch lines.[14] It is obvious that the HB possessing only two Bloch lines is the simplest HB. The configuration in Fig. 1 is just a coarse model to portray the primary character of HB. In order to obtain their accurate distribution in a nanoscale HB, the magnetization were numerically calculated by the object oriented micro-magnetic framework (OOMMF) software based on solving the Landau-Lifshitz-Gilbert (LLG) equation.[15] The generated magnetization configuration were input to a home-made program to compute the induced magnetic field in a larger space containing the specimen and then to simulate the LTEM Fresnel images under different tilting conditions and optical defocuses. The experimental images of NiMnGa alloy were acquired in JEOL LM-2100F and Tecnai F20 electron microscope with Lorentz lens. A home-made TIE plugin written in Digital Micrograph script was used to recover the exit electron wave phase $\varphi(x,y)$ from the simulated or the experimental images and to deduce the magnetization configuration from the recovered phase.

The object for calculation in OOMMF is a $(Mn_{1-x}Ni_x)_{65}Ga_{35}$ (x=0.45) slab of dimension $192 \times 192 \times 130$ nm with mesh $1.5 \times 1.5 \times 2$ nm. The saturation magnetization $M_S$ is $8 \times 10^5$ A/m$^3$, the magneto-crystalline anisotropy constant $K_u$ is $2.65 \times 10^5$ J/m$^3$ and the exchange constant A is $1.4 \times 10^{-11}$ J/m, respectively. Two arcs (Fig. 1) were set as the initial spin configuration and then the system relaxed to stable state. The final results shown in Fig. 2 indicates that the magnetization pattern in the middle $x$-$y$ plane of the nanobubble is similar to the model in Fig. 1b but the distribution on two surfaces change remarkably. That is reasonable because the magnetostatic energy should decrease if the magnetization over there are parallel to the surface.

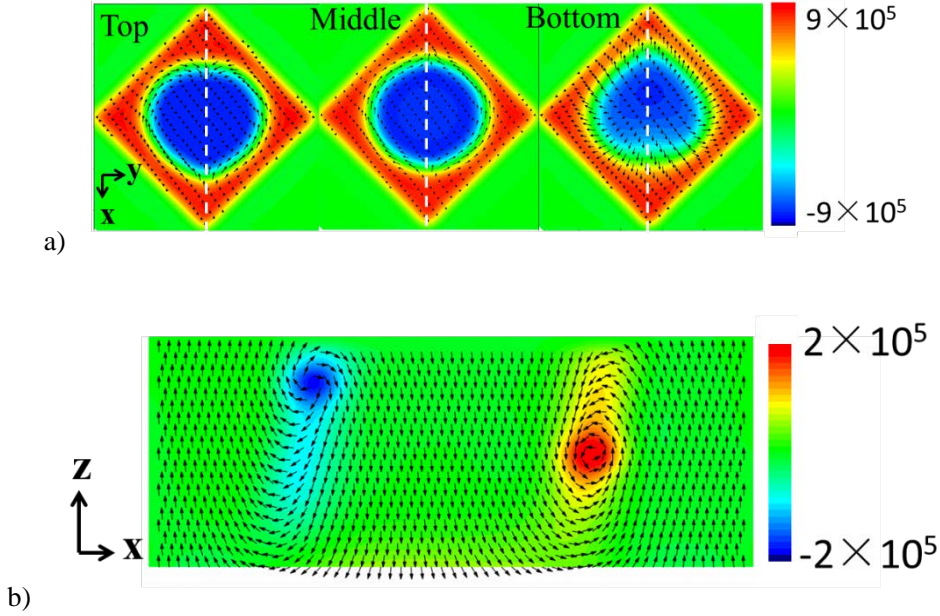

a)

b)

Fig. 2 Magnetization distribution in the sections of a hard nanobubble. a) At $z$ = 65 nm, 0 nm and -65 nm. b) At $y$ = 0 nm. The length and orientation of the arrows indicate the amplitude and direction of in-plane components while the color visualizes the strength of the out-of-plane part. The image sizes are 278 × 278 nm in a) and 278 × 115 nm in b), respectively.

Fig.3 shows the simulated defocus LTEM images of the nanobubble without tilting. The two arcs can be distinguished clearly and each arc contains a white-black stripe in the same sequence, implying the same magnetization orientation there. After the TIE processing, the reconstructed $M'_x$ and $M'_y$ are displayed in the first row of Fig. 4a and 4b, denoted with their $q_0$. To verify the influence of specimen tilting, the results for different tilting angles are also displayed in Fig. 4a and Fig. 4b, corresponding two tilting axes, respectively. In the following figures, the color, with the assistance of the direction of the arrows, illuminates the orientation of the recovered magnetization $M'$ while the length of the arrows represents the relative magnitude of $M'$, which are more interesting for the magnetic structure characterization. The retrieved $M'_x$ and $M'_y$ are similar to the real $M_x$ and $M_y$ when $q_0$ is small (0.1 pixel or 2.6×$10^{-4}$ nm$^{-1}$). As $q_0$ is 1 pixel or 2.6×$10^{-3}$ nm$^{-1}$, the opposite magnetization appears in the core of nanobubble and a typical bi-spiral or bi-skyrmion emerges. For a larger $q_0$ (5 pixel or 1.3×$10^{-2}$ nm$^{-1}$), new arcs appear in the core and periphery of the nanobubble. Specimen tilting leads to the distortion in final features. Similar behavior can be found in the experimental survey for NiMnGa alloys, as shown in Fig. 5 where just the effect of $q_0$ is investigated. These results illustrate that the processing parameter $q_0$ and specimen placement in LTEM can complicate the recovered magnetic configuration and bring the confusion in interpreting the actual magnetic structures in the samples.

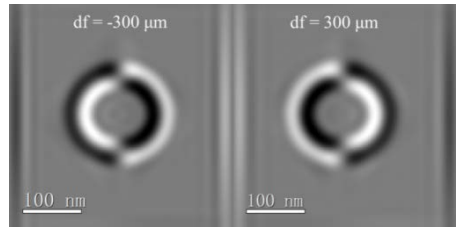

Fig. 3 The simulated LTEM images of a hard nanobubble at different defocuses. (Accelerate voltage: 200 keV, Cs: 5 m, Cc: 100 Å)

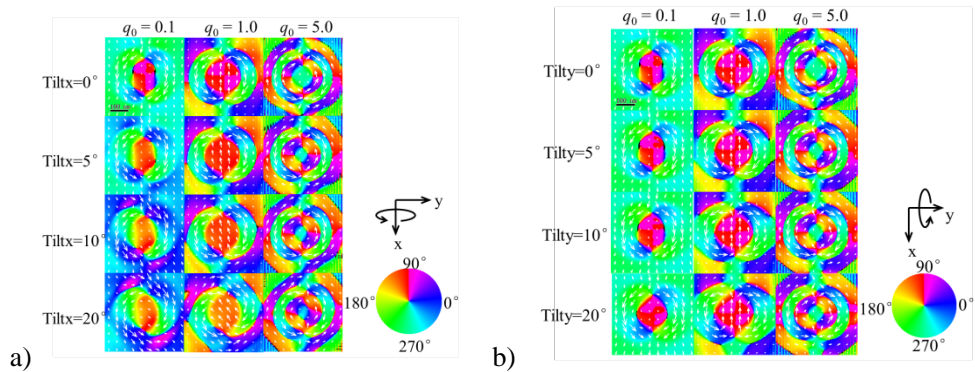

Fig. 4 The influence of $q_0$ and the sample tilting around different axes, retrieved with TIE processing (Unit of $q_0$ is pixel, 1 pixel = $2.6 \times 10^{-3}$ nm$^{-1}$).

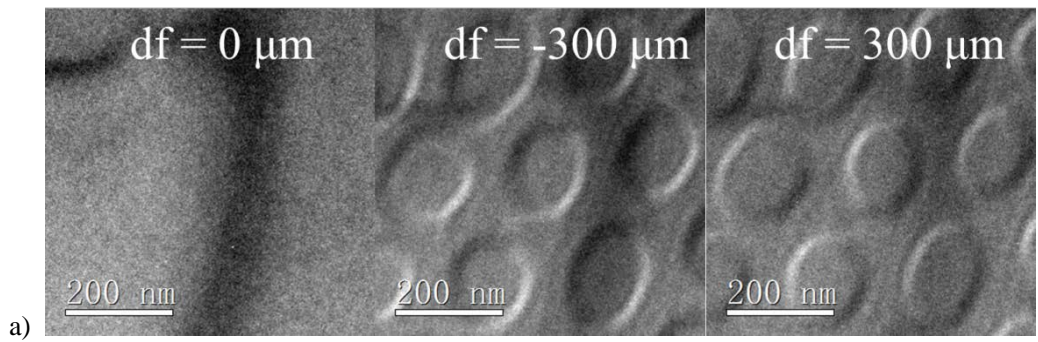

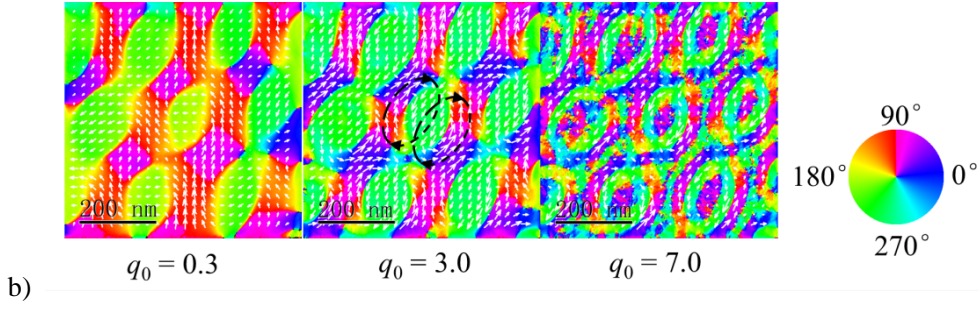

b)

| $q_0 = 0.3$ | $q_0 = 3.0$ | $q_0 = 7.0$ |
|---|---|---|

Fig. 5 a) The LTEM images of NiMnGa and b) the retrieved magnetic configuration with different $q_0$ (Unit of $q_0$ is pixel, 1 pixel = $1.6 \times 10^{-3}$ nm$^{-1}$).

The reproduction of bi-skyrmions in the LTEM images of a traditional magnetic hard nanobubble may provide a hint to explain the observed similar phenomena in some center-symmetric magnetic materials where DMI does not exist. LTEM images exhibit the projection characteristics of the magnetic induction $\vec{B}$ in the space where electrons penetrate, not $\vec{M}$ in the specimen, although $\vec{B}$ is much stronger in the specimen than in free space. So the retrieved phase images may mask the real magnetization configuration, especially for a thin specimen which owns a considerable portion of the magnetization parallel to the surfaces. When the sample varies its orientation relative to the electron beam, the situation becomes complex. Moreover, the magnetic structure recovered with FT assistant TIE method is sensitive to the parameter $q_0$. In eq. 2', $q_0$ restrains the low frequency noise or removes the influence of the diffraction contrast which always forms the background in the LTEM images, such as those dark stripes on the images in Fig. 5a and Fig. 6a. Unfortunately, the low frequency signal is lost at the same time, which is illustrated in Fig. 4 and Fig. 5. On the other hand, much higher $q_0$ may highlight the high frequency noise shown in Fig. 6b which demonstrates these influences on the experimental images. In practice, a reliable $q_0$ can be estimated from Eq. 2'. If the loss of low frequency information at $q_{min}$ is no more than 10%, the max $q_0^{max}$ should be $q_0^{max} \approx \sqrt{2}q_{min}/4.3 \approx q_{min}/3$.[4, 5] For instance, if $q_{min} = 5 \times 10^{-3}$ nm$^{-1}$ is needed to guarantee the fidelity of a 200 nm-sized disk, $q_0$ should be smaller than $1.6 \times 10^{-3}$ nm$^{-1}$. It can be seen that most bi-skyrmions features emerge when $q_0$ is beyond this critical value in Fig. 4 - Fig. 6.

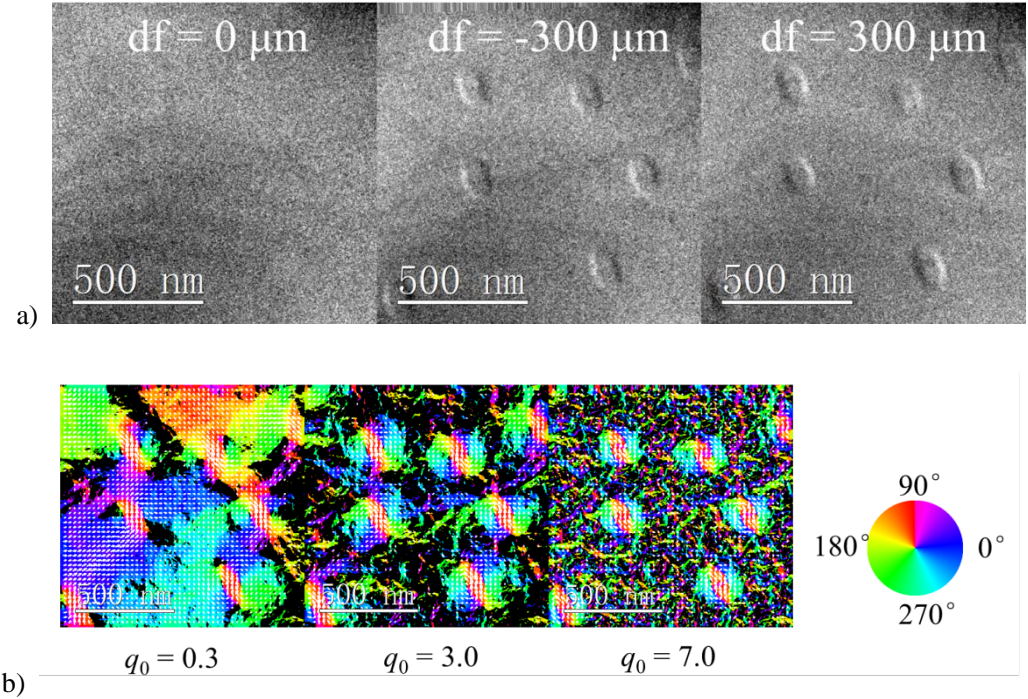

a)

b)      $q_0 = 0.3$              $q_0 = 3.0$              $q_0 = 7.0$

Fig. 6 a) The LTEM images of NiMnGa and b) the retrieved magnetic configuration with different $q_0$ to demonstrate the influence of low frequency contrast and high frequency noise (Unit of $q_0$ is pixel, 1 pixel = $8\times10^{-4}$ nm$^{-1}$).

The contrast-performance in LTEM images of hard nanobubbles is different from a pure Neel spiral domain wall. Bi-skyrmion contrasts could appear in a pure Neel spiral domain, if its vertical axis is tilted away from the electron beam in LTEM.[13] However, that feature will vanish when its axis is along the electron beams. So, a hard nanobubble always contributes image contrast in LTEM whatever the sample direction is, as shown in Fig. 3. Thus, it may be an effective method to identify what is the actual magnetic structure behind the bi-skyrmions features by tilting the sample in a wide angle range.

In summary, the magnetization and LTEM images of a magnetic hard nanobubble have been simulated to investigate the origin of the bi-skyrmions in centrosymmetric magnetic materials. Systematic studies on both the simulated and experimental images indicate that the retrieved magnetic configuration are sensitive to the parameter $q_0$ of TIE approach and $q_0$ can modify the magnetic appearance easily. Meanwhile, sample tilting also deteriorates the distortion. The reproduction of bi-skyrmion feature by adjusting $q_0$ hints that a simple magnetic structure can lead to a complex phenomenon when an improper parameter is applied to deal with the LTEM images. As a comparison, electron holography (EH) in LTEM should be a recommended method to avoid those artificial features since that acquires the images under in-focus condition and does not require the parameter such as $q_0$ in TIE processing when recovering the exit-wave phase. However, a special biprism and the additional procedures to remove the mean inner potential of

**Supplementary Material**

The supplementary material includes the simulation configuration in OOMMF, the noise influence on image processing, the experimental initialization of the hard nanobubbles and the contrast variation when the sample was tilted in experiment.

**Acknowledgment**

This work was supported by the Ministry of Science and Technology of the People's Republic of China (Grant Nos. 2016YFA0202504, 2017YFA0206200 and 2017YFA0303202) and National Natural Science Foundation of China (Grant 11874410).

Magnetic hard nanobubble: a possible magnetization structure behind the bi-skyrmion

Yuan Yao (姚湲)[1,†,*], Bei Ding (丁贝)[1,2,†], Jie Cui (崔婕)[1,2], Xi Shen(沈希)[1], Yanguo Wang (王岩国)[1], Wenhong Wang (王文洪)[1], Richeng Yu (禹日成)[1,2]

1. Beijing National Laboratory of Condensed Matter Physics,
Institute of Physics, Chinese Academy of Sciences,
Beijing 100190, China
2. School of Physical Sciences, University of Chinese Academy of Science
Beijing 100049, China

[†] Equal contribution for this work.
[*] Corresponding author, email: yaoyuan@iphy.ac.cn

## Supplemental Materials

### 1. OOMMF simulation

Two methods were used to generate the magnetic configuration of hard nanobubble. One was to set two arches spin structures as the initial spin configuration and relax the system. Another is to apply in-plane magnetic field to transform a normal Bloch soft bubble into a hard nanobubble and then remove the in-plane magnetic field to obtain a relax state. The chirality of the hard bubbles created by those two methods are same. The former is used to simulate the LTEM images.

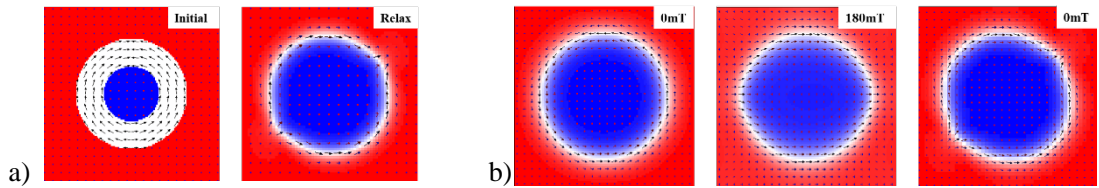

Fig. s1 Detail of the magnetization distribution at Z=0nm. a) We construct two arches spin structures as the initial state and relax the system. b) Based on a normal Bloch soft bubble, we apply in-plane magnetic field to obtain a hard nanobubble and then remove the in-plane magnetic field to relax the system.

### 2. The noise influence

The experimental images have been acquired in JEOL LM-2100F and F20 equipped Lorentz lens, respectively. The beam was near parallel and the exposure time was 3s to get a good contrast and S/N ratio. The noise was below 10% of the signal, estimated from the flat contrast area of the specimen in the in-focus images. So the short noise about 10% of the image intensity has been added in the simulated images to investigate the influence of the experimental noise.

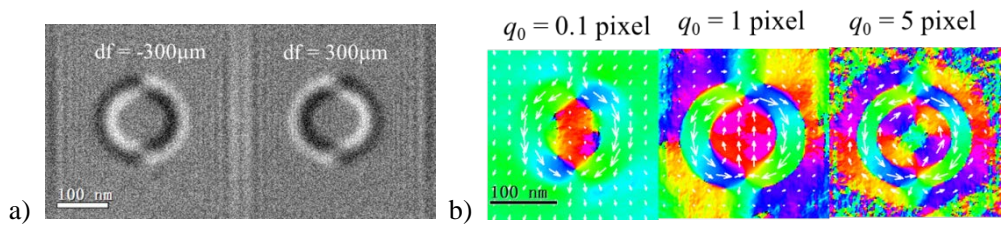

Fig. s2 a) The simulated image with short noise. b) The retrieved magnetization maps. All parameters for simulation and TIE processing are as same as in Fig. 3 and Fig. 4 for untilted specimen.

## 3. Initial process of the bubbles

An appropriate field cooling procedure was applied on the [001] thin plate of the $(Mn_{1-x}Ni_x)_{65}Ga_{35}$ (x=0.45) alloy. A magnetic field perpendicular to the plate surface could lead to the soft bubbles. When the sample was tilted little, the hard nanobubbles were observed.

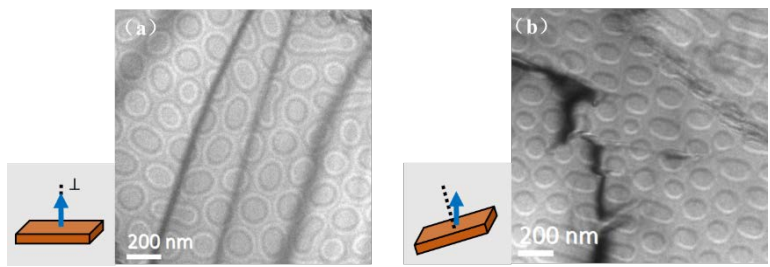

Fig. s3 a) The soft bubbles were induced by the perpendicular field. b) The hard bubbles appeared when sample was tilt little.

## 4. Judgement of the type of the biskyrmion

The specimen in LTEM experiments has been tilted from -10° to 10° to investigate whether the features vary with the tilting angles. As the rotation angle changes from -10° to 10°, hard bubbles still accompanied with a pair of Bloch line, which was different from the Neel skyrmion.

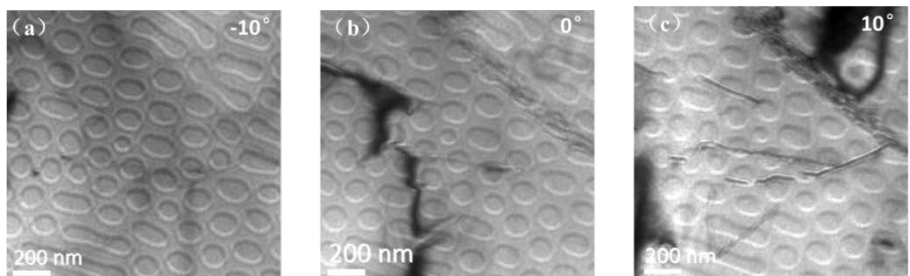

Fig. s4 The experimental defocus images of a NiMnGa alloy tilted in LTEM.

## 5. The contrast oscillation in the domain wall

In the simulated images in Fig. 3, there are the contrast oscillations in each domain wall. For the experimental images in Fig. 5 and Fig. 6, such oscillations appear weaker. That difference may come from the sample thickness or imaging parameters, such as defocus. An image for a thicker sample shown below demonstrates the similar contrast vibrations as same as in the simulated image.

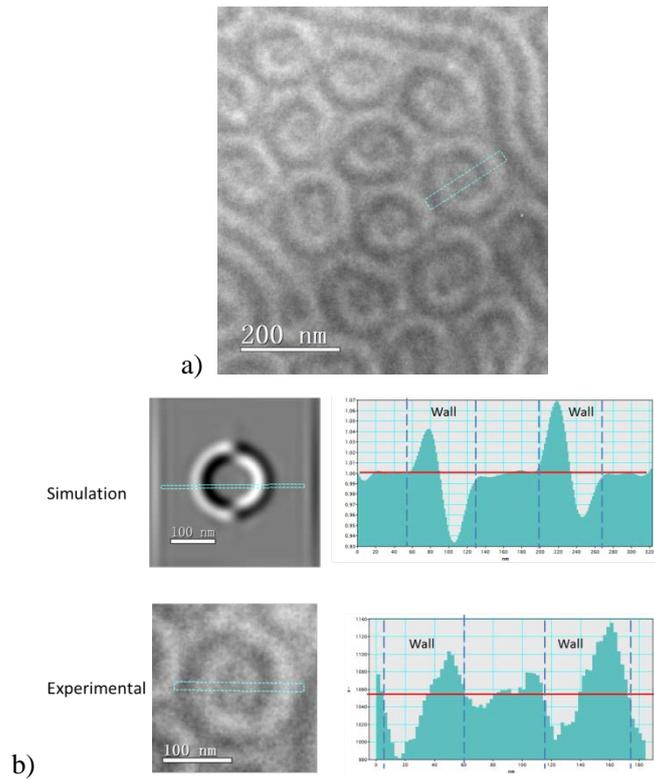

Fig. s5 a) An experimental image of a thicker NiMnGa alloy. b) The contrast oscillations in the domain walls of the simulation image and the experimental image, respectively.